\documentclass[structabstract]{aa}  
\usepackage{natbib}
\bibpunct{(}{)}{;}{a}{}{,}
\usepackage{graphicx}
\usepackage{txfonts}
\begin{document}
   \title{Deep R-band counts of z$\approx$3 Lyman break galaxy candidates with the LBT. $\thanks{Observations were carried out using the Large Binocular Telescope at Mt. Graham, AZ. The LBT is an international collaboration among institutions in 
the United States, Italy, and Germany. LBT Corporation partners are The University of Arizona on behalf of the Arizona university system; Istituto Nazionale 
di Astrofisica, Italy; LBT Beteiligungsgesellschaft, Germany, representing the Max-Planck Society, the Astrophysical Institute Potsdam, and Heidelberg University; 
The Ohio State University; and The Research Corporation, on behalf of The University of Notre Dame, University of Minnesota, and University of Virginia.}$ $\thanks{Appendix Tables A1, A2, 
and A3 are only available in electronic form at the CDS via anonymous ftp to cdsarc.u-strasbg.fr (130.79.128.5) or via http://cdsweb.u-strasbg.fr/cgi-bin/qcat?J/A+A/vol/page.}$}

   \author{K. Boutsia \inst{1}, A. Grazian \inst{1}, E. Giallongo \inst{1}, M. Castellano \inst{1}, L. Pentericci \inst{1}, 
   A. Fontana \inst{1}, F. Fiore \inst{1}, S. Gallozzi \inst{1}, F. Cusano \inst{2}, D. Paris \inst{1}, R. Speziali \inst{1}, V. Testa \inst{1}}

   \institute{INAF - Osservatorio Astronomico di Roma, Via Frascati 33, I - 00040 Monteporzio (RM), Italy\\
   INAF - Osservatorio Astronomico di Bologna, Via Ranzani 1, I - 40127, Bologna, Italy\\
                           \email{konstantina.boutsia@oa-roma.inaf.it}}

  \authorrunning{Boutsia, K. et al.} 
  \titlerunning{Deep counts of z$\approx$3 LBG candidates}

 
  \abstract
   {}
   {We present a deep multiwavelength imaging survey ($UGR$) in 3 different fields, Q0933, Q1623, and COSMOS, 
   for a total area of $\sim$1500arcmin$^2$. The data were obtained with the Large Binocular Camera on the Large Binocular Telescope. }
   {To select our Lyman break galaxy (LBG) candidates, we adopted the well established and widely used color-selection criterion (U-G vs. G-R).
   One of the main advantages of our survey is that it has a wider dynamic color range for U-dropout selection than in previous studies. This 
   allows us to fully exploit the depth of our R-band images, obtaining a robust sample with few interlopers. In addition, for 2 of our fields we have 
   spectroscopic redshift information that is needed to better estimate the completeness of our sample and interloper fraction.  }
   {Our limiting magnitudes reach 27.0(AB) in the R band (5$\sigma$) and 28.6(AB) in the U band (1$\sigma$). This dataset was used to 
   derive LBG candidates at z$\approx$3. We obtained a catalog with a total of 12264 sources down to the 50\% completeness
   magnitude limit in the R band for each field. We find a surface density of $\sim$3 LBG candidates arcmin$^{-2}$ down to R=25.5, where completeness is $\ge$95\% 
   for all 3 fields. This number is higher than the original studies, but consistent with more recent samples.}
   {}


   \maketitle
%

\section{Introduction}

Lyman-break galaxies (LBGs) are star-forming galaxies that emit very little flux in the observed
UV when they are at redshifts higher than z=2.5. This is because the stellar radiation
with energy beyond the Lyman limit (912\AA) is absorbed by the surrounding neutral hydrogen and by 
the intervening neutral clouds between the galaxy and the observer. Thus the SEDs of these galaxies are characterized by a sharp drop 
at wavelengths shorter than the 912\AA~rest frame \citep{Mad95} and by a steep increase between the 912\AA~and the 1216\AA~rest frame. 
Such features have been used extensively during past decades to create substantial samples of LBGs at high redshifts. 
More specifically, the filters $UGR$ have been used for selecting U dropouts 
that are candidate LBGs at z$\approx$3 \citep[e.g.,][]{Steidel96,Giava02,Steidel03,Capak04,Sawicki05,Noni09}.
\citet{Steidel03} applied this method to 17 high Galactic-latitude fields and presented a sample of 2347 photometrically selected LBG
candidates down to a magnitude limit of 25.5 in the R band, corresponding to $\sim$1500$\AA$ rest frame at z$\approx$3, in an area of $\sim$3200arcmin$^2$. 
After a spectroscopic follow-up \citep{Steid04}, the success rate for LBGs at redshift
z$\sim$3 was on the order of 78\%. Thus the adopted color selection provides samples with low contamination that can be used 
for spectroscopic follow up to do statistical analyses of the LBG population and to study the physical properties 
of LBGs, such as stellar masses and the UV slope. A statistically significant LBG sample, associated with deep U-band imaging, 
can also be used to derive stringent upper limits for the escape fraction of UV ionizing radiation from LBGs. 
In addition, such a dataset is suitable for studying clustering by applying a two-point correlation function analysis, as well as 
for deriving the fraction of AGNs embedded in such galaxies by combining it with X-ray observations.

After the first effort by \citet{Steidel03}, similar surveys have been conducted that reach different magnitude limits. \citet{Sawicki05} covered a 
relatively small area (169arcmin$^2$), but reached a deeper magnitude limit of R=27.0 (50\% point sources detected). More recently, \citet{Raf09}
presented a sample of LBGs at z$\sim$3, using both photometric redshifts and color selection. Their color-selection criterion uses a filter set that is slightly different 
(u-V vs. V-z) from the one established by \citet{Steidel03}  (U-G vs. G-R), and their sample is complete up to V$\approx$27.0, which corresponds 
to R$\approx$26.5 for this type of sources. 
\citet{Noni09} present deep imaging in the GOODS area (630arcmin$^2$), with 50\% completeness in LBG selection at R$\approx$26.0. \citet{ly11} used Subaru images, covering an area of 
870arcmin$^2$, with 5$\sigma$ limiting magnitudes of R=27.3, but limited the search for LBG candidates at R=25.5. An extended survey 
has been presented by \citet{VDBurg10}, who used data from the Deep CFHT survey, which covers 4~sq. deg and reaches R=27.9 at 5$\sigma$, 
although their U-dropout number counts only seem to be complete up to R=26.0. The most recent and extended survey is the one conducted by \citet{Bian13}, 
which covers 9~sq. deg in the NOAO Bo$\ddot{o}$tes Field, although it is rather shallow, selecting LBG candidates down to R=25.0.
 
Because of the variety of instruments and filters used to select these LBG candidates at z$\sim$3, the selection biases in the various samples are difficult 
to quantify and at times they lead to diverging results. For example, \citet{LeFev05} find that the number density of galaxies between z=1.4 and z=5 is 1.6 to 6.2 times higher 
than earlier estimates based mainly on the work of \citet{Steid04}. Such discrepancies in the number density also lead to discrepancies in the 
derived LFs \citep[e.g.,][]{Iwat07,Sawic06,Reddy08}. 

We used the Large Binocular Camera \citep[LBC,][]{gial08} at the Large Binocular Telescope (LBT) to obtain a multiwavelength dataset ($UGRIZ$) on
three different fields to derive a new sample of LBG candidates at z$\sim$3 through photometric selection. One of the main advantages of our 
survey is that we have spectroscopic redshifts for two fields (Q0933 and Q1623) and accurate photometric redshifts for COSMOS. In this third field (COSMOS), 
spectroscopic redshifts are also available, but in a different redshift range than the one we are targeting in this study. These are useful, nonetheless, 
since we can use them to assess the interloper fraction of our selected candidates. Thus, this is one of the few surveys that combine deep data in a large area 
with spectroscopic data, giving us a direct way of assessing the completeness and contamination of our sample. 

The LBC is a wide field binocular imager which gives us the opportunity to probe large areas with deep imaging, particularly in the U band, 
where it is extremely efficient. In fact, the total area covered by our survey is $\sim$1500arcmin$^2$.
Moreover, LBC also includes a custom-made U-band filter, U$_{Special}$, that is particularly efficient and centered on bluer wavelengths ($\lambda_{central}$=355nm), 
making it more suitable for selecting LBG candidates compared to standard U band. According to the standard color-selection criterion, established by \citet{Steidel03}, 
for an average color of G-R=0.5, LBG candidates should have U-R$\ge$2.1. This means that for selecting LBG candidates, at R$\leq$26.5 we need a 1$\sigma$ magnitude 
limit in U band of at least 28.6, in order to exploit the full dynamic color range. For fainter R-band magnitudes, the candidates could show up as upper limits because 
of incompleteness effects in the U band and not because of their intrinsic SED. 

The main purpose of this work is to present the full catalog of LBG candidates, selected in the three fields, down to a 50\% completeness magnitude limit (R=26.1-27.0). 
This catalog will serve as the database for future works that will estimate the UV slope and stellar mass of LBG candidates, especially in the COSMOS field where 
additional photometry is available. 
The presented U-band magnitudes can be used to improve photometric redshift estimates in the COSMOS field. The brightest part of this sample can be used for spectroscopic 
follow-up, and this new spectroscopic sample would help refine the color-selection criterion for LBGs further. Part of this dataset has already been used by \citet{Grazian09} 
to present deep U-band counts in the Q0933 field, while the extended dataset was used by \citet{Bout11} to derive a stringent upper limit to the escape 
fraction of ionizing photons of LBGs at z$\sim$3.3. By adding new spectroscopically confirmed candidates, an even more accurate estimate of the escape fraction 
could be obtained based on this sample.

In the following we focus on the number counts of galaxies in the $UGR$ bands and the counts of LBG candidates at z$\sim$3 in the R band, selected using 
the traditional color-color criteria, along with the slopes derived by the double power law fit of the galaxy number counts. 
More precisely, in Section 2 we describe how the imaging data were obtained. In Section 3 we present the multiband photometry and the galaxy counts. In Section 4, 
we present the selection criteria for deriving the LBG candidates and the number counts. In Section 5 we discuss completeness and the effect of interlopers, while in 
Section 6 we summarize our results. Throughout the paper we adopt the AB magnitude system.  
 

\section{Imaging data}

The dataset was obtained with the LBC \citep{gial08} at the LBT on Mount Graham in Arizona. The LBC is a double
imager installed at the prime focus of each of the two 8.4m mirrors mounted on
the telescope. Each camera has an unvignetted field of view of
$23\arcmin\times23\arcmin$ with a sampling of 0.226 arcsec/pixel. Each channel is
optimized in a different wavelength range. The LBC-Blue, is optimized in the UV-B range
and the red channel (LBC-Red) is optimized in the $VRIZY$ wavebands. Both cameras have an eight-filter
wheel, for a total of 13 available filters covering all wavelengths from the
ultraviolet to the near infrared. 

We repeatedly observed three fields in five bands ($UGRIZ$), with the last images obtained in 2010, resulting in a very deep multiband imaging dataset. 
In this work we present counts in the three bands, $UGR$, that were used to derive the LBG candidates. We have actually used two similar U-band 
filters (U$_{bessel}$ and the custom made U$_{special}$), as well as the R filters mounted on both channels (R-Sloan$_{blue}$ and R-Sloan$_{red}$), 
which have slightly different throughput. Thus, a total of five filters were used to collect this dataset and all transmission curves are presented in Fig. 1.

\begin{figure}
 \centering
   \includegraphics[width=8cm]{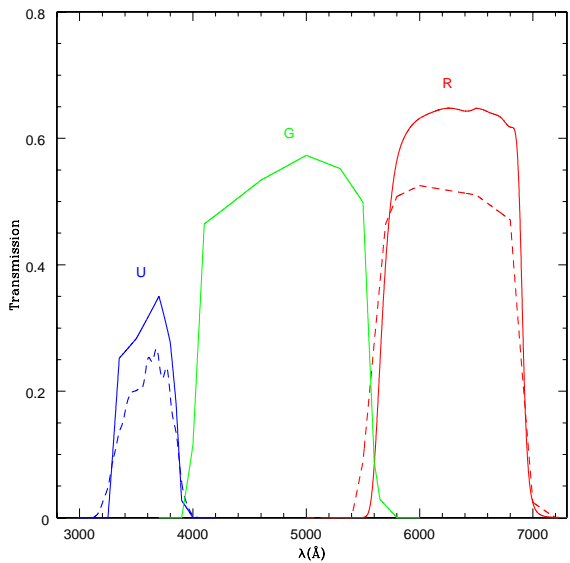}
      \caption{Transmission of LBC filters used in our survey. These curves include the telescope and instrument response but not the atmospheric response, since 
      the data have been obtained during several observing runs with varying weather conditions. For the U and R filters 
      there are two curves that represent the different filters. The blue solid line represents U$_{special}$ and the blue dashed line represents U$_{bessel}$.
      The red solid line represents R-Sloan in the red channel, and the dashed line is the transmission of the R-Sloan filter in the blue channel. 
      There is only one G-Sloan filter on LBC-Blue.}
     \label{filters}
\end{figure}

Two of the fields, Q0933 and Q1623, are centered on bright QSOs and are part of the 
Steidel dataset \citep{Steidel03,Reddy08}. The third pointing covers an area of the wider 
COSMOS field \citep{scov07,lill07}, where extensive multiwavelength data is already available, 
as well as ongoing spectroscopic follow-up \citep{Lill09}. The data was reduced using the LBC pipeline 
and applying standard techniques for imaging data: bias-subtraction, flat-fielding,
sky-subtraction and final co-addition after the appropriate astrometric
corrections were applied. A detailed description of the procedure can be found in \cite{gial08}. 
The LBC pipeline also produces a map of the standard deviation for each scientific 
image, directly from the raw science frame, as described in detail by \cite{Bout11}.
These rms maps are used to calculate upper limits in the bands where the sources are not detected. 

The average exposure times for single frames have changed during the campaign (average values: 300-400sec in U band and 120-160sec in the G and R bands). 
Seeing, magnitude limits, total exposure time, and filter sets are different from field to field. 
The two Steidel fields only include one LBC pointing in the 
U band each (with dithering applied), while for the COSMOS field we used two LBC pointings with 
an average exposure time of 2.2hr (7900sec) and 5.8hr 
(20800sec), respectively. These two fields, overlap for an area of 352arcmin$^2$, 
where the total exposure time reaches about eight hours in the U band (28700sec), and this is 
practically 70\% of the final effective area we use for COSMOS. The total 
effective area for the three fields is 1465.4 arcmin$^{2}$ (502arcmin$^{2}$ in COSMOS, 505arcmin$^{2}$ in Q1623, 458.4arcmin$^{2}$ in Q0933). 
In Table 1 we show the average properties of all final coadded mosaics in each field. 

\begin{table*}
\caption{\label{t1}LBC fields}         
\centering                   
\begin{tabular}{cccccc}     
\hline                 
FIELD & Coordinates& Filter & Exp.Time (s.) & Seeing ($\arcsec$)& Mag. limit(AB) \tablefootmark{1}\\    
\hline                        
Q0933  & 09:33:35.739 +28:39:50.64 & U$_{BESSEL}$ & 27300 & 1.04 & 26.19\\     
Q0933  &  & G$_{SLOAN}$  & 14900 & 0.99 & 26.42\\
Q0933  &  & R$_{SLOAN}$  &  9000 & 1.02 & 25.33\\ 
Q1623  & 16:25:44.125 +26:47:07.71& U$_{SPECIAL}$& 30200 & 0.99 & 26.17\\ 
Q1623  & & G$_{SLOAN}$  & 15000 & 0.99 & 26.68\\ 
Q1623  & & R$_{SLOAN}$  & 16700 & 0.84 & 26.38\\ 
COSMOS \tablefootmark{2} & 09:59:55.855 +02:12:26.98 & U$_{SPECIAL}$& 28700 & 0.95 & 26.07\\
COSMOS & & G$_{SLOAN}$  & 12200 & 1.02 & 26.48\\
COSMOS & & R$_{SLOAN}$  & 12000 & 0.93 & 25.71\\
\hline      
\end{tabular}
\tablefoot{
\tablefoottext{1}{Magnitude limit is at 10$\sigma$ and in a 2$\times$fwhm aperture for Q0933 and Q1623 and 3$\times$fwhm in COSMOS as explained in section 3.}\\
\tablefoottext{2}{The exposure time corresponds to the overlapping deepest area of the mosaic, while the mag. lim. and seeing are average values.} 
}
\end{table*}

\section{Multiband photometric catalogs and galaxy counts}

We used the R-band mosaic as detection image in each field and ran 
Source Extractor \citep{Bertin96} in dual mode for computing the magnitudes in the U and G filters.
Two important parameters for the depth and completeness of our catalogs are the threshold
adopted for the detection of sources (DETECT\_THRESH) and the minimum area of associated pixels 
above this threshold (DETECT\_MINAREA). Their values depend strongly on the average seeing of the final
mosaics, and the threshold has been fixed to 3/sqrt(minarea) with 
minimum areas of 16 (Q0933), 11 (Q1623), and 13 pixels (COSMOS). 
To account for the seeing differences, we used different photometric apertures for each 
image, which correspond to the area of a circle with a diameter of 2$\times$fwhm, where fwhm is the full-width at half-maximum of 
bright, unsaturated stars in the source image.
It is known that at faint magnitudes, the mag$_{AUTO}$ underestimates the total photometry of 
a source, especially for those with extended morphology. For this reason, 
for sources with an ISOAREA corresponding to a disk diameter of less than 2$\times$fwhm, we used corrected aperture magnitudes. 
The total R magnitude calculated for each source after the aperture correction is defined as  

\begin{equation}
R_{TOT} = R_{APER} + C_{APER} + C_{GAL},
\end{equation}
where
\begin{equation}
C_{APER} =<mag_{AUTO} - mag_{APER}>_{stellar} 
\end{equation}
\begin{equation}
C_{GAL} =<\Delta m_{stellar}-\Delta m_{gal}>_{R_{band}}.
\end{equation}

The corrections were calculated as follows. First, we took the average 
$<$mag$_{AUTO}$ - mag$_{APER}$$>_{stellar}$ value for bright stellar sources. 
Such a correction should compensate for flux losses due to aperture size.
This initial aperture correction does not take the loss of flux for more extended galaxies into account. 
Thus, to account for sources with extended morphologies we
calculated a further correction based on the morphology of the object, by obtaining
aperture photometry of the sources in the R-band image using apertures with diameters 2$\times$fwhm and 3$\times$fwhm. Then we calculated the quantity 
$\Delta m_{stellar}$=mag$_{APER_{3\times fwhm}}$-mag$_{APER_{2\times fwhm}}$ for sources with stellar morphologies
(Class.Star$>$0.95) and the same quantity for source with extended morphologies, $\Delta m_{gal}$. To correct 
for the morphology, leaving the colors untouched, we applied this additional 
average quantity, $<\Delta m_{stellar}-\Delta m_{gal}>$, to the stellar 
aperture correction obtaining eq.(1).  

In the COSMOS field we combined two pointings observed under different 
conditions, and the seeing in our final mosaic is different between the overlapping region 
and the region of lower exposure time. For this reason, we computed photometry in apertures of 3$\times$fwhm.
In this case, we only applied the stellar aperture correction, and there was no additional $\Delta m$ quantity 
calculated for the extended sources: 

R$_{TOT_{COSMOS}}$= $R_{APER}$ + $<$mag$_{AUTO}$ - mag$_{APER_{3\times fwhm}}$$>$$_{stellar}$.\\

To test the reliability of our catalogs further, we estimated the contamination by false
detections using the negative image technique \citep[e.g.][]{Dick04,Yan04}. This consists in
applying the same detection parameters as used for extracting the sources on the negatives of
the detection images, after subtracting all known objects. In all our fields, the contamination
by spurious sources is below 7\% for R$<$26.0.
We then compared the number of objects in each magnitude bin for our fields
with the counts of sources in the R band calculated by \citet{Metc01} and \citet{Capak04} (see
Fig.2). The distributions appear consistent in all fields, which indicates that our photometry
is well calibrated. 

\begin{figure}
 \centering
   \includegraphics[width=8cm]{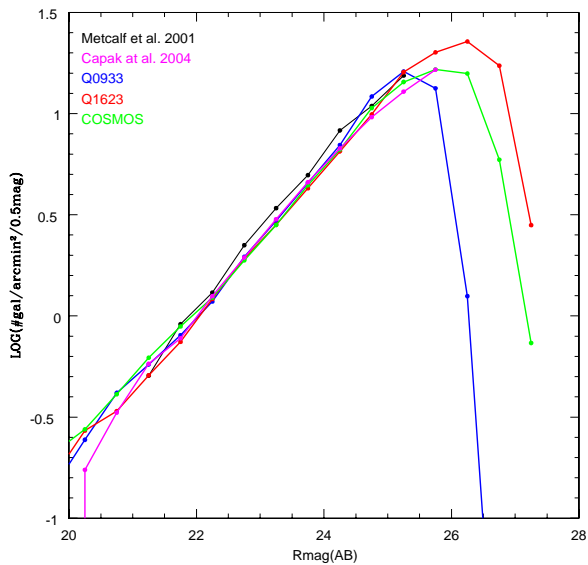}
      \caption{Galaxy counts (number of sources per 0.5mag per arcmin$^{2}$) in the R band derived from our analysis separately in each field.
      There is good agreement with previous estimates \citep{Metc01,Capak04}, which is a strong indication of the robustness of our photometry.}
     \label{COUNTSR}
\end{figure}

Photometry in the other bands was obtained by running SExtractor in dual mode, using the mosaics in the R band as the detection image. 
To correctly account for color terms, the magnitude of the sources in the other bands is calculated as

\begin{equation}
mag_{filter,corr}=R_{TOT}-(mag_{aper,R}-mag_{aper,filter}).
\end{equation}

We derive the counts of only the extended sources by applying a cut 
in the fwhm. More precisely, by plotting the histogram of the fwhm for 
all sources up to a certain relatively bright magnitude (ranging from 23.0 to 24.5 in the R band, 
depending on the depth of the image), we distinguish one tight peak, representing the 
fwhm distribution of the stellar sources, followed by a second smoother distribution that corresponds 
to the extended sources. We select as extended sources, those that have a
fwhm larger than the value where the two distributions meet. 
This cut occurs at 1.3", 1.01", and 1.13" for Q0933, Q1623, and COSMOS, respectively.

In Table 2 we present the number counts of galaxies (in 0.5mag bins) for the three fields in our dataset. The galaxy counts were obtained by detecting the sources 
directly in the G and U bands. Thus, the magnitude in this case was not calculated using eq. (4), but by applying the same photometric method as for the 
R band (i.e., eq. (1)-(3) for each band). In Fig.3 we show the counts in the G band and in Fig.4 the counts in the U band for the whole galaxy sample. In both 
figures we see that our galaxy counts are consistent with all previous studies, which is a further indication of the robustness of our photometry. 
In particular for the G band, we compared our results to the work of \citet{Shim06} that presented deep G counts using the MegaCam on the CFHT. Their 
catalog is 50\% complete down to G=26.5, but they only present number counts down to G=25.0. We see that their data are fairly consistent with our curves, 
considering field-to-field variations. In the U band we compared our counts with those of \citet{Capak04}, \citet{EliMor06}, \citet{Rov09}, and \citet{Grazian09}. 
Again, our results are in agreement with all the aforementioned surveys, and the field-to-field variations are relatively small at U$\geq$22.0.   

 \begin{figure}
 \centering
   \includegraphics[width=8cm]{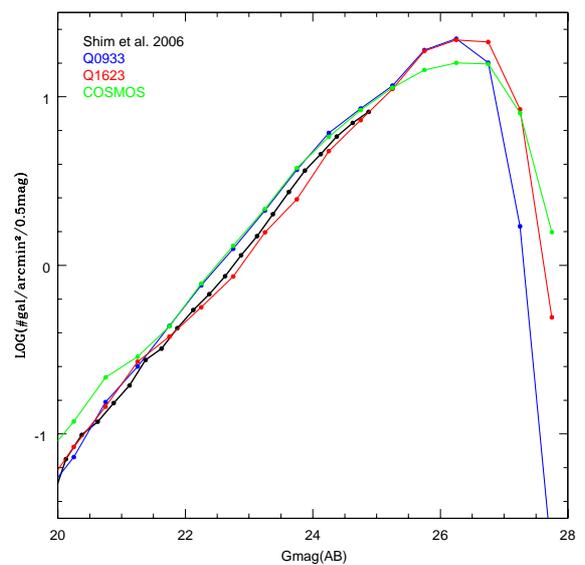}
      \caption{ Galaxy counts in the G band for each of our fields separately. The magnitudes of the galaxies for this plot were derived by single-mode photometry, 
      detecting sources directly in the G-band images. Although we see some scatter in the bright end, there is overall agreement when compared to the 
      literature \citep{Shim06}.}
     \label{COUNTSG}
\end{figure}

\begin{figure}
 \centering
   \includegraphics[width=8cm]{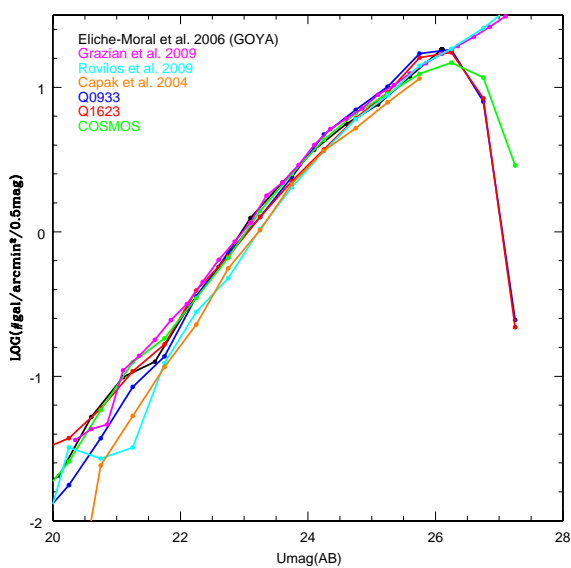}
      \caption{ Number counts of galaxies per 0.5mag per arcmin$^{2}$ in the U band for the fields in our survey, which have been derived by single-mode photometry. 
      We see that there is very good agreement between our estimates and previous studies \citep{Capak04, EliMor06, Rov09,Grazian09}.}
     \label{COUNTSU}
\end{figure}

\begin{table*}
\caption{Galaxy number counts in 0.5mag bin in the $UGR$ bands.}         
\centering                   
\begin{tabular}{c|ccc|ccc|ccc}     
\hline                 
filter & & R & & & G & & & U & \\
\hline
mag. bin & Q0933  & Q1623 & COSMOS & Q0933 & Q1623 & COSMOS & Q0933 & Q1623 & COSMOS \\    
\hline                        
 20.0 - 20.5 &  112  &   137  &  138 &    40 &    45  &    69 &    9  &   20  &   15\\ 
 20.5 - 21.0 &  191  &   171  &  206 &    85 &    78  &   126 &   19  &   32  &   34\\ 
 21.0 - 21.5 &  264  &   257  &  312 &   138 &   144  &   167 &   43  &   58  &   73\\ 
 21.5 - 22.0 &  368  &   376  &  445 &   240 &   203  &   253 &   70  &   89  &  107\\ 
 22.0 - 22.5 &  540  &   612  &  618 &   418 &   303  &   454 &  182  &  210  &  203\\ 
 22.5 - 23.0 &  898  &   970  &  945 &   688 &   461  &   761 &  361  &  362  &  386\\ 
 23.0 - 23.5 & 1356  &  1426  & 1413 &  1163 &   844  &  1259 &  646  &  677  &  814\\ 
 23.5 - 24.0 & 2082  &  2160  & 2222 &  2022 &  1325  &  2201 & 1221  & 1195  & 1469\\ 
 24.0 - 24.5 & 3207  &  3289  & 3334 &  3348 &  2557  &  3368 & 2402  & 1986  & 2533\\ 
 24.5 - 25.0 & 5572  &  5006  & 5353 &  4682 &  3902  &  4855 & 3557  & 3267  & 3854\\ 
 25.0 - 25.5 & 7401  &  8100  & 7199 &  6396 &  5984  &  6601 & 5140  & 4939  & 5273\\ 
 25.5 - 26.0 & 6115  & 10141  & 8290 & 10389 & 10029  &  8382 & 8722  & 8568  & 7218\\ 
 26.0 - 26.5 &  574  & 11474  & 7918 & 12123 & 11681  &  9250 & 9278  & 9325  & 8629\\ 
 26.5 - 27.0 &    3  &  8715  & 2971 &  8739 & 11347  &  9102 & 4065  & 4482  & 6801\\ 
 27.0 - 27.5 &    0  &  1419  &  369 &   936 &  4523  &  4656 &  125  &  117  & 1678\\ 
\hline 
area (arcmin$^2$) &  458.4 & 505 & 502 & 549 & 537 & 582 & 509.6 & 537 & 583 \\
\hline
\end{tabular}
\end{table*}
 

\section{LBG candidates' selection criteria}

For selecting LBG candidates at redshift $z\sim3$, we use the (U-G) vs. (G-R) color-color 
selection technique that was introduced by Steidel and collaborators. 
The LBC filters are slightly different from the filter set used by Steidel and for this
reason we modified their color selection to fit our filter set using the relation found in \citet{gial08}.
This relation has been checked with synthetic color predictions derived from galaxy spectral synthesis models \citep{bc03}. 
The optimized cuts for our filter set are

\begin{equation}
U-G >= 1.20\times(G-R)+0.96 
\end{equation}
\begin{equation}
G-R >= -0.3
\end{equation}
\begin{equation}
G-R <= 1.0.
\end{equation}

We verified our cuts by plotting the sources with known spectroscopic redshift
on the color-color diagram. As expected most of the sources with redshift above 2.7 are 
within our cuts (80\% for Q0933 and 94\% for Q1623). On the 
other hand, those sources that have been selected as photometric LBG candidates at z$\sim$3 
by \citet{Steidel03} but then turned out to be at lower redshift are mostly outside 
our cuts (93\% for Q1623 and 62.5\% for Q0933). This means that our sample
suffers less from contamination by lower redshift galaxies than previous studies, and this is mainly due
to our deeper images in the U band that allow us to select our LBG candidates better.
In Fig.5 we show the (U-G) vs. (G-R) diagram for the Q1623 field, which is the deepest area studied in this work, 
along with the sources with spectroscopic redshifts (Reddy, private communication).   
In the COSMOS field, where only photometric redshifts are available in this redshift range \citep{lill07}, we find that 
for magnitudes 22.0$<$R$<$24.5, 76\% of the sources with 2.7$\leq z_{phot} \leq$3.4 are within our cuts 
(thus selected as LBG candidates), while only 8\% of sources with lower photometric redshift (2.0$<z_{phot}<$2.7) 
contaminate our sample.

The choice of adapting our color cuts to the Steidel criterion was made to be able to 
compare our results directly with previous studies. In fact, according to stellar models \citep{Pick98,Gunn83}, in the lower 
right region delimited by our cuts, there could be contamination by stellar sources that are difficult 
to purge at faint magnitudes by morphological criteria alone. By strictly following the limits of the synthetic SEDs for LBGs, the color cut should 
be steeper, leaving out the stellar locus. But such a cut would also leave out several spectroscopically 
confirmed LBGs that do not follow the model tracks exactly.  

In Fig.6  we show the results of the LBG counts derived with our selection criteria in the three different fields and show how these counts
compare to the Steidel results \citep{Steidel03}, in the fields we have in common. In the Q0933 field, our results are in very good agreement with the
Steidel counts at bright magnitudes and begin to diverge at around R$\sim$25, where our curve becomes steeper, since we find 
more LBG candidates. This is because our 10$\sigma$ magnitude limit is deeper than their adopted 3$\sigma$ R-band magnitude cut, 
thus our detections suffer less by incompleteness at the faint end. The LBG counts in Q0933 alone, as already noted by \citet{Steidel03}, are steeper
than the average counts derived after combining the entire dataset. In Q1623, a field that is very close to the galactic plane, it is rather
difficult to exclude all stars. In fact, we see that the Steidel counts for this field are heavily contaminated by stars at all
magnitudes brighter than 24.5. Our morphological criterion for separating extended sources seems to produce acceptable results down
to a magnitude of 23.0 in the R band. In the brightest magnitude bin (22.0-23.0) however, ours too is contaminated by stars, and this bump is 
visible in our combined-counts curve. At the faint end, our LBG counts for Q1623 are still a lot steeper than the results presented by Steidel. 
This is our deepest field in the R band, reaching one magnitude fainter than previous studies. The counts in the COSMOS field are consistent 
with our results in the other two fields. Our combined-counts curve is consistent with previous studies in the bright end and becomes steeper for R$>$24.0. 

\begin{figure}
 \centering
   \includegraphics[width=8cm]{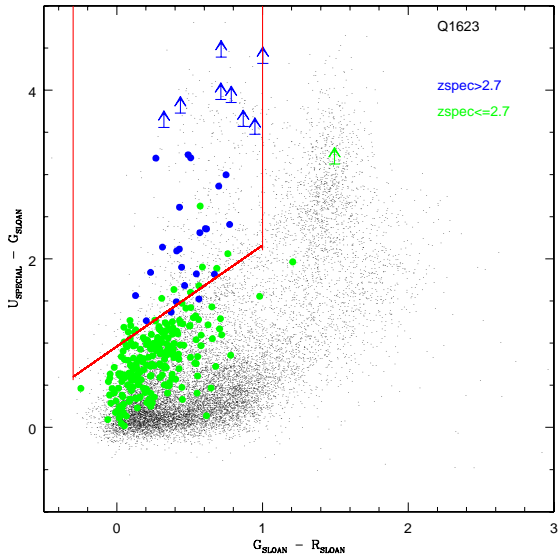}
      \caption{ The U-G vs. G-R diagram that was used for the selection of the LBG candidates in the field Q1623, 
      that is the deepest of the three. Red lines indicate the selection color locus. 
      Green symbols indicate sources with spectroscopic redshift $z_{sp}\leq2.7$, and blue symbols show LBGs with $z_{sp}>2.7$ 
      (Reddy, private communication). Vertical arrows indicate upper limits in the U band. Black points indicate all sources detected 
      in the field with R$\leq$25.0. Sources with spectroscopic redshift have magnitudes R$\leq$25.5.}
     \label{UGR}
\end{figure}

In Fig.7 we show the counts of our LBG candidates in the R band and compare them to other works.
Our curve is higher than the original Steidel counts \citep{Steidel99}, but this is a common trend 
in all subsequent similar works. Actually, \citet{Capak04} used a slightly different filter set (U-B vs. B-R) to select
the U dropouts, and their R counts, after removing the stars, reach a magnitude limit of $<$26.0, which is almost 0.5 magnitude 
brighter than our deepest R-band image. This is also true for their U-band images, since their 5$\sigma$ limit corresponds to our
10$\sigma$ magnitude limit. Thus the images used by \citet{Capak04}, are 0.5 magnitude deeper than Steidel's and are consistent with our counts at the bright end. 
At the faint end, they are more consistent with \citet{Reddy08}, but that their R images are 0.5 magnitude brighter than our 
dataset could explain this discrepancy. In fact, recent works presented by \citet{Noni09} and \citet{ly11}, are in good agreement 
with our results. Although consistent with the latter studies at the faint end, the counts presented by \citet{VDBurg10}  
appear significantly lower with respect to all previous surveys for R magnitude brighter than 24.0. This could be due to the different wavelength coverage of the U filter 
adopted in the CFHT survey, which is offset to redder wavelengths with respect to standard U filters. Consequently, the redshift selection 
function could be centered at a slightly higher redshift than previous studies, where there are fewer objects. According to \citet{VDBurg10}, 
the estimated mean redshift for their U dropouts is $<z>$=3.1$\pm$0.1, while for the LBC filter set, it is $<z>$=2.9. 
The most recent survey, presented by \citet{Bian13}, also shows counts that at the bright end are systematically lower than the work of \citet{Reddy08}. 
As seen in Fig.7, our counts are between the two curves reported by \citet{Reddy08} and \citet{Bian13} in the bright magnitude range, and this could either 
be due to cosmic variance or to slight differences in the color criteria used in each survey. 

In Table 3 we present the number counts for our LBG candidates in bins of 0.5mag. 
The last column presents the average number counts per arcmin$^2$, weighted with the area and corrected for completeness (see par. 5). 
For magnitudes fainter than 25.5, each field contributes to this average, up to the magnitude limit where relative completeness is $>$80\%. 

The total surface density of LBG candidates for each field in the magnitude range 23.0$<$R$<$25.5, weighted by area, 
is 3.23 sources/arcmin$^{2}$ in COSMOS, 3.73 sources/arcmin$^{2}$ in Q0933, and 2.49 sources/arcmin$^{2}$ in Q1623. 
The average number for our total survey is 3.15$\pm$0.62 sources per arcmin$^{2}$, which leads to a cosmic variance of $\sim$20\%. 
Using the original catalogs presented by \citet{Steidel03}, in the fields we have in common, we find a surface density of LBG candidates 
of 2.5 sources/arcmin$^{2}$ for Q0933 and 1.8 sources/arcmin$^{2}$ for Q1623. We obtain this result without taking into consideration subsequent adjustments for
completeness and interlopers based on the spectroscopic follow-up. According to \citet{Steidel03}, the overall surface density in their survey is 1.8 LBG 
candidates per arcmin$^{2}$, uncorrected for completeness, which is lower than the estimate based on our analysis by 40\%. This could be explained by the 
fact that this magnitude (R=25.5) corresponds to their 3$\sigma$ limit, so their
survey could be less complete. In more recent surveys, \citet{Rov09} find 2.3 sources per arcmin$^{2}$ and \citet{Raf09} find 3.7$\pm$0.6 (LRD field) and 
4.3$\pm$0.2 (KDF field) for V$<26.0$, which corresponds to our magnitude limit of R$<25.5$. For fainter magnitudes, \citet{Sawic06} find 8.8 LBG
candidates/arcmin$^2$ (50\% completeness at R$\approx$27.0) and \citet{Noni09} report a surface density of 7.3 arcmin$^{-2}$ at a similar magnitude limit, although they 
are already $>$50\% incomplete by R=26.0. In our deepest field, Q1623, where we also have 50\% completeness at R=27.0, we find 10.8 LBG candidates/arcmin$^{2}$, 
based on the raw number counts. Although it is difficult to directly compare with previous surveys, due to different completeness fractions at that magnitude limit, 
we see that our surface density is consistent with previous surveys at faint magnitudes, when taking the cosmic variance into account.

In Appendix A we provide the LBG candidate catalog in each field, down to the 50\% completeness magnitude limit in the R band. 
The magnitudes presented in these catalogs have been calculated using the R band as the detection image, and for the other two bands (G and U), we report 
the corrected aperture magnitudes, derived in dual-mode photometry according to eq. (4). For each magnitude we also report the associated error.
Negative magnitude values in the G and U bands, indicate upper limits at 1$\sigma$. For the R band, where the sources were detected, we also present a S/N estimate, 
the measured fwhm in pixels, and the morphological class attributed to the source by SExtractor, with values closer to zero indicating an extended morphology.

\begin{table}
\caption{Counts of LBG candidates for each field in the R band }         
\hspace{-0.5cm}
\begin{tabular}{ccccc}     
\hline                 
mag & Q0933 & Q1623 & COSMOS & ALL \tablefootmark{1}\\    
\hline                        
 23.0 - 23.5 &  19 &   16 &   26 & 0.04 \\ 
 23.5 - 24.0 &  43 &   53 &   71 & 0.11 \\ 
 24.0 - 24.5 & 179 &  139 &  201 & 0.34 \\ 
 24.5 - 25.0 & 504 &  331 &  458 & 0.84 \\ 
 25.0 - 25.5 & 965 &  719 &  864 & 1.71 \\ 
 25.5 - 26.0 & 965 & 1235 & 1191 & 2.45 \\ 
 26.0 - 26.5 & 110 & 1657 & 1257 & 3.38 \\ 
 26.5 - 27.0 &   0 & 1301 &  456 & 3.77 \\ 
 27.0 - 27.5 &   0 &  195 &   42 &  --- \\ 
\hline				   
area (arcmin$^2$)& 458.4 & 505 & 502 & \\
mag.lim. at 50\% compl.& 26.1 & 27.0 & 26.5 & \\
total cand to compl. lim. & 2744 & 5451 & 4069 & \\
\hline 
\end{tabular}
\tablefoot{
\tablefoottext{1}{Average number of LBG candidates per arcmin$^2$, weighted for area and corrected for incompleteness}
}
\end{table}

\begin{figure}
 \centering
   \includegraphics[width=8cm]{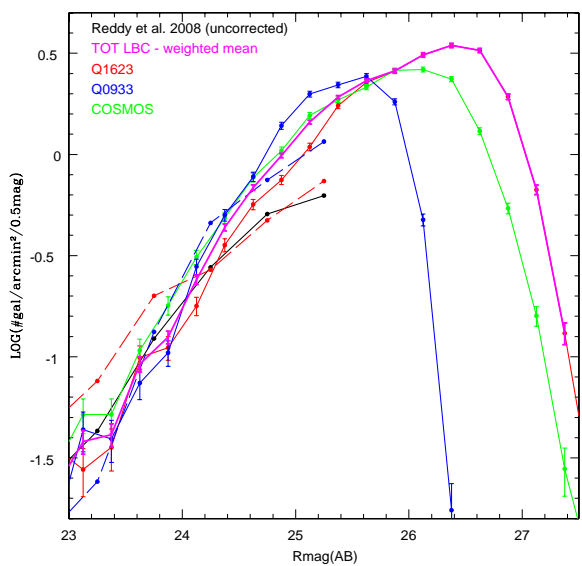}
      \caption{ Number counts of LBG candidates in each field of our survey. The dashed lines represent the number counts of the candidates in the relative fields 
      obtained by Steidel's catalogs \citep{Steidel03}. The total counts for all Steidel fields were obtained from Table 2 of \citet{Reddy08}, and they are not 
      corrected for contamination or completeness. The magenta curve shows the weighted mean of the total raw LBG counts in this survey. }
     \label{countsUGR}
\end{figure}

\begin{figure}
 \centering
    \includegraphics[width=8cm]{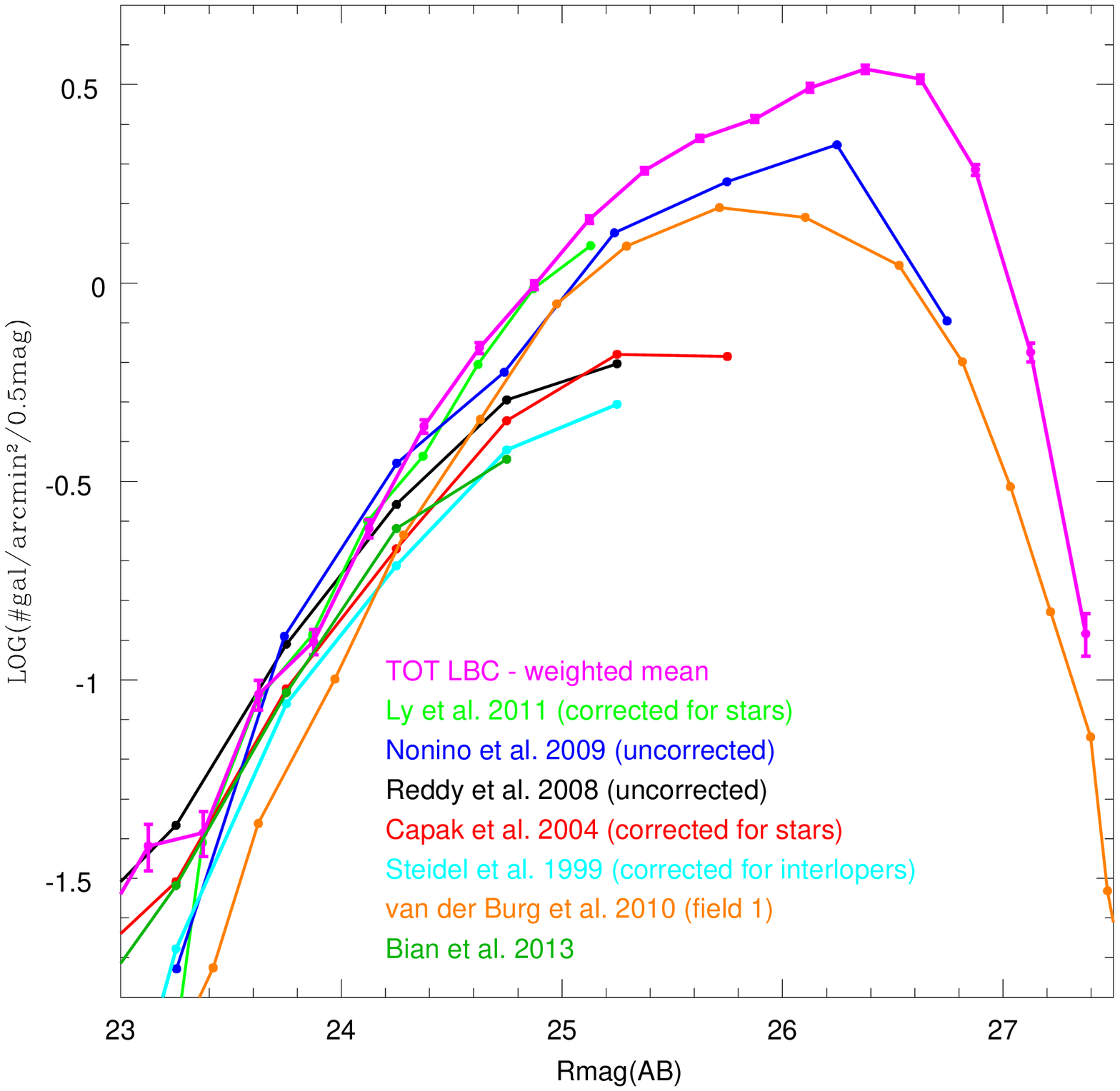}
      \caption{ Weighted mean of the uncorrected LBG counts of our survey. The LBG count numbers by \citet{Noni09} and \citet{Reddy08} are not corrected for contamination. 
      The count numbers by \citet{ly11} and \citet{Capak04} are corrected for star contamination. We also show the original number-count curve by \citet{Steidel99}, 
      corrected for intelopers. Among recent works, we present the curve for field-1 by \citet{VDBurg10} and the surface density presented by 
      \citet{Bian13} for the Bo$\ddot{o}$tes survey.}  
     \label{countsUGRn}
\end{figure}

\section{Completeness and interlopers}

To estimate the completeness of our sample, we use a set of simulations. First, we select stellar sources 
with magnitudes in the range of 23.0$<$R$<$24.0 and fwhm$<$fwhm$_{cut}$ (see section 3), and we stack them to create a point spread function (PSF) template. 
We then randomly place 1000 of these stacked PSFs at different R magnitudes in our images. We then repeat the detection procedure and check the percentage of
sources recovered by our method. We apply the same method, also using galaxy profiles. To create the galaxy profiles, we selected LBG candidates 
in three different magnitude bins centered at 24.0, 25.0, and 26.0, after visual inspection for contamination by neighboring sources. We made three stacked images, 
one for each magnitude bin, which we then randomly placed and recovered, as described above.
We show an example of these simulations in field Q1623 and how the various completeness curves compare in Fig.8. 
We see that using a PSF, which is created by stars, gives very similar results to the profiles created by stacking 
faint galaxies (R=26.0) and they correspond to our lower limit in the completeness curves. By using 
profiles based on brighter galaxies (R=25.0 and 24.0), the corrections are larger, since brighter galaxies are more extended and this 
could compromise the detection efficiency \citep{coh03}. 
Since the stellar PSF gives us the lower limit for the completeness curve, and it is also very similar to the faintest galaxy curve, we
decided to repeat the simulations using only stellar PSFs for the other two fields, as a conservative approach.

Another concern in our survey is the fraction of lower redshift interlopers contaminating the LBG candidate selection. To quantify the number of interlopers 
in our sample, we use the spectroscopic data available in COSMOS for z$<$1.5 and in the Q1623 field for z$>$1.5. 
To perform simulations in the magnitude range 25.0$\leq$R$\leq$27.0 we selected fairly bright sources (23.0<R<24.50) 
for a total of 240 objects with known redshift. The sample is divided into redshift bins of 0.2, and the 
number of real sources in our spectroscopic catalogs has been normalized according to the redshift distribution obtained at faint 
magnitudes (in bins of 0.5mag) by the Millennium simulation \citep{Kitz07}.

This technique allows us to reproduce the actual redshift distributions in different magnitude ranges starting from our spectroscopic catalogs
which are limited at bright R-band magnitudes. Starting from the observed $UGR$ photometry of the spectroscopic catalog, renormalized in number, 
we extrapolate the photometry to fainter magnitudes, reproducing the observed magnitude-error relations for each band.
All sources with spectroscopic redshift used in the input catalog are correctly allocated in the color-color diagram; thus, 
all sources inside the color-color selection locus are LBGs at z$\approx$3, and all sources outside have lower redshifts. 
From the output catalog we can assess the number of sources that are detected as bright LBG candidates but at fainter magnitudes 
have moved outside the color-color selection locus, because of the photometric error. 

We can also estimate the number of sources that 
were outside the color-color locus and are now detected as spurious LBG candidates.  
To repeat the simulations for the brighter magnitude range of 24.0$\leq$R$\leq$25.0, the same method was adopted after 
restricting the input catalog of sources with known redshift to 23.0$\leq$R<24.0 (155 objects). 
We see that the maximum of the net effect of the scatter for sources going in and out of the color locus (net contamination) is reached around R=26.0, 
where we have $\sim$21\% more sources with z<2.7 entering the color-color selection area. 
For fainter magnitudes, according to the redshift distributions, there are less sources of lower redshift that can end up in the 
color-color locus of LBGs because of photometric error, so it is reasonable that the percentage of contamination is actually lower. 
Our results are in accordance with the contamination reported by \citet{hilde07}, who find a 20\% contamination 
in the magnitude range 22.0<R<26.0. 

To quantify the contamination by stars, we select all stars in the COSMOS field based on a BzK selection and on 
spectroscopic data (using publicly available photometry).  After cross-correlating this catalog to the catalog of our 
LBG candidates in COSMOS, we only find ten sources in common out of $\sim$4000 candidates, which are evenly distributed in the 
magnitude range 23.0<R<26.3. Thus, our contamination by stars is negligible (0.25\%).

In Table 4 we present the completeness calculated for each field, in bins of 0.25mag, as well as the "net contamination" fraction for the same magnitude bins, after a spline fitting 
of the values obtained through simulations. In Fig.9 we show the relative plot with the detection efficiencies, where the "net contamination" fraction is also presented,  
although we do not correct our number counts for this effect.
We see that down to a magnitude of R=25.5, completeness is above 99.5\% (practically complete) for all our fields. For Q0933, completeness rapidly decreases
after this limit, reaching 50\% at a magnitude of R=26.1, making it the shallowest of the three fields. The COSMOS field follows, since it is 50\% complete 
at R=26.5, and Q1623 reaches this limit at R=27.0, leading to the deepest in our survey. Actually, Q1623 is the field with the deepest photometry in all 
three bands, as shown in Table 1, and with the greatest number of spectroscopically confirmed LBGs, thus it is ideal for exploring the limits of our color selection. 

\begin{figure}
 \centering
   \includegraphics[width=8cm]{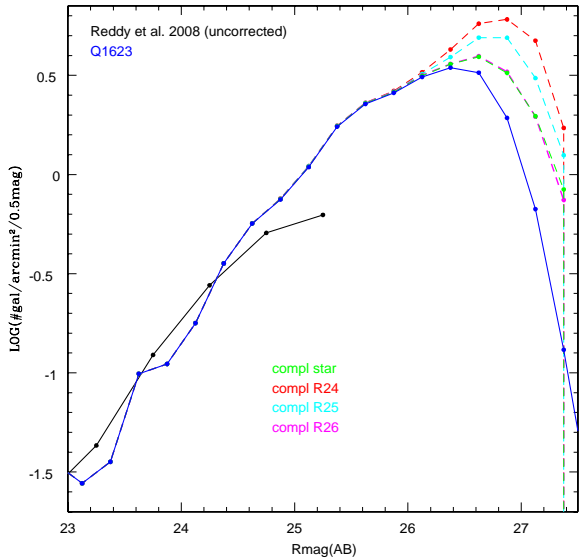}
      \caption{Completeness corrections derived from simulations in Q1623, which is the deepest of our fields, using PSFs created by stacking
      real stellar and extended sources in different bins of magnitude. }
     \label{simulQ16}
\end{figure}

\begin{table}
\caption{Completeness in the R band for each field and total "net contamination" fraction.}         
\centering                   
\begin{tabular}{ccccc}     
\hline                 
mag & Q0933 & Q1623 & COSMOS & net contam. (\%)\\    
\hline                        
    23.125 & 1.000 & 1.000 & 1.000 & ---\\ 
    23.375 & 1.000 & 1.000 & 1.000 & ---\\ 
    23.625 & 1.000 & 1.000 & 1.000 & ---\\ 
    23.875 & 1.000 & 1.000 & 1.000 & ---\\ 
    24.125 & 1.000 & 1.000 & 1.000 &  3.95\\ 
    24.375 & 1.000 & 1.000 & 1.000 &  6.67\\ 
    24.625 & 1.000 & 1.000 & 1.000 & 11.17\\ 
    24.875 & 1.000 & 0.996 & 1.000 & 13.76\\ 
    25.125 & 0.999 & 0.995 & 1.000 & 13.73\\ 
    25.375 & 0.995 & 0.996 & 0.996 & 13.51\\ 
    25.625 & 0.952 & 0.993 & 0.995 & 14.77\\ 
    25.875 & 0.745 & 0.993 & 0.984 & 18.14\\ 
    26.125 & 0.385 & 0.988 & 0.928 & 20.77\\ 
    26.375 & 0.117 & 0.960 & 0.672 & 19.74\\ 
    26.625 & 0.068 & 0.832 & 0.304 & 14.97\\ 
    26.875 & 0.035 & 0.593 & 0.128 &  5.99\\ 
    27.125 & 0.017 & 0.340 & 0.070 & ---\\ 
    27.375 & 0.015 & 0.155 & 0.021 & ---\\ 
    27.625 & 0.000 & 0.000 & 0.000 & ---\\ 
\hline 
\end{tabular}
\end{table}

\begin{figure}
 \centering
   \includegraphics[width=8cm]{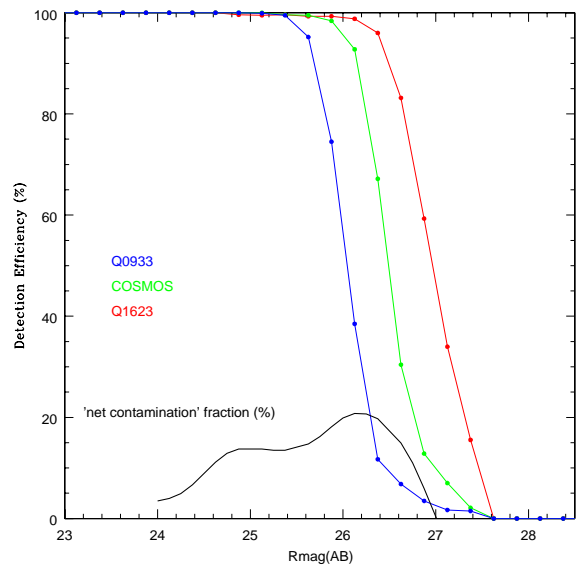}
      \caption{Detection efficiencies in the R band derived by simulations in all 3 fields using PSFs created by stacking stellar sources down to 24.0 in the R band.
      The black curve indicates the percentage of the "net contamination" fraction we expect in the total sample as a function of magnitude.}
     \label{eff}
\end{figure}

Based on these curves we calculate a weighted average for the counts of LBG candidates in all our fields, corrected for completeness. 
To account for the different depths of our fields, we used the counts of the entire dataset up to mag=25.5 in R band. After this
magnitude limit, completeness for the Q0933 field is flattening and the corrections become increasingly large (completeness $\sim$95\% already at R=25.6), 
so we only use the counts of the other two fields, Q1623 and COSMOS, for the faint bins of our survey(25.5$\leq$R$<$26.2). The last bins (26.2$\leq$R$\leq$27.0) 
are entirely computed using only Q1623. We then fit this curve with a double power law, in order to derive the slope. To be conservative, 
we apply the fit only down to R=26.6, where completeness is $>$80\%. After this magnitude limit, we see that the curve is flattening even 
though it is corrected for completeness. 
The double power law is described by the following formula:
\begin{equation}
N = \frac{N^{*}}{10^{-\alpha*(m-m^{*})}+10^{-\beta*(m-m^{*})}}.
\end{equation}

In Fig.10 we show both the curve corresponding to the weighted average corrected for completeness and the double power law fit.  
The two slopes derived from this fit are $\alpha$=1.04 and $\beta$=0.13, with the break located at m*=25.01 and $\log$N*=0.39,
where N is the number of galaxies per arcmin$^2$ per 0.5mag bins. 
The fit for the lower limit curve at 1$\sigma$ is $\alpha$=1.16, $\beta$=0.20, m*=24.8, $\log$N*=0.25, and for the upper limit curve the fit corresponds to
$\alpha$=0.94, $\beta$=0.06, m*=25.2, and $\log$N*=0.51. There are no records of similar analysis in the literature, so we cannot directly 
compare our results with previous studies. 
 
\begin{figure}
 \centering
    \includegraphics[width=8cm]{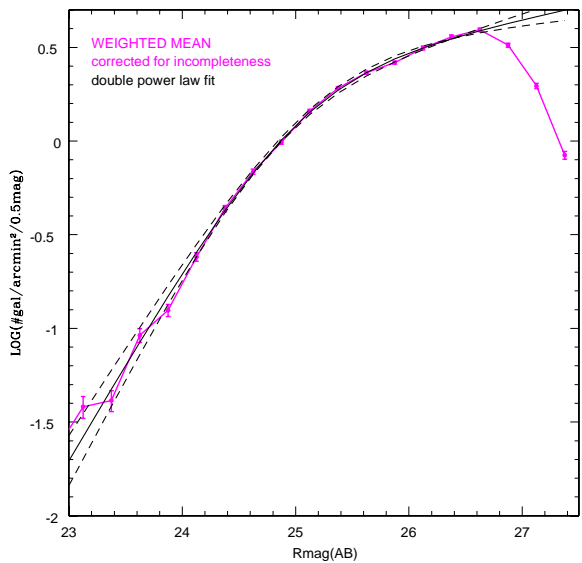}
      \caption{ Weighted average for the LBG counts, after correcting for completeness. The average is based on all 3 fields until
      R=25.5 and, for fainter magnitudes, only the two deepest fields are considered (Q1623 and COSMOS), with a conservative cut 
      made at $\sim$80\% completeness. We also show the double power law fit of this curve and the derived 
      values for $\alpha$=1.04 and $\beta$=0.13, m*=25.01 and $\log$N*=0.39, where N is the number of galaxies per arcmin$^2$ per 0.5mag bins. 
      The dotted lines show the upper and lower limit curves.}
     \label{pendALL}
\end{figure}


\section{Discussion and summary}
We presented a deep multiband imaging survey with the LBC, covering an area of $\sim$1500arcmin$^2$.
We reobserved two fields used in Steidel's original survey \citep[Q0933 and Q1623,][]{Steidel03}, where we obtained deeper R- and U-band imaging. 
A similar dataset was also obtained for the COSMOS field, where there is available public multiband photometry as well as photometric and spectroscopic redshifts.
We reached 50\% completeness at R magnitude of 27.0, 26.1, and 26.5 for Q1623, Q0933, and COSMOS, respectively.
The 1$\sigma$ magnitude limit in the U band is between 28.5-28.7 on the whole area, which is a good compromise between depth and total area, 
compared to other surveys, so far. A significant advantage of our sample is that the U band is much deeper than previous samples. 
For a limiting magnitude of R=27.0 (50\% completeness) in our deepest field and an average magnitude 
in the U band of 28.6 at 1$\sigma$ (2$\times$fwhm apertures), this is the only survey with a wide dynamic 
range in the color selection, allowing us to robustly select LBG candidates with minimum contamination. In comparison, the CFHT survey \citep{VDBurg10}, 
although reaching 27.9 in the R band (5$\sigma$ for point sources), is shallower in the U band than ours.
At the bright end we are fairly consistent with the new survey presented by \citet{Bian13} in the NOAO Bo$\ddot{o}$tes Field, 
but the two surveys start diverging after R$>$24.0, since the latter one is two magnitudes shallower than our survey, 
although it is covering a much larger area ($\sim$9~sq. deg). 

Comparing our candidates with existing spectroscopy in the Steidel fields, where spectroscopic redshifts are available, we show that the deeper U-band 
dataset allows us to better separate confirmed LBGs at z$\approx$3 from lower redshift interlopers. 
Although we have less contamination by low-redshift sources, we can see in Fig. 8 that the slope of our LBG counts is actually steeper than previous studies, 
suggesting that there are more LBGs at faint magnitudes. The slopes we derived are $\alpha$=1.04 at the bright end and $\beta$=0.13 at the faint end, 
with a break at m*=25.01 and $\log$N*=0.39. We find an average surface density of 3.15 LBG candidates per arcmin$^2$ down to R=25.5, which rises to 10.8 LBG 
candidates per arcmin$^2$ when we go as faint as R=27.0.

This dataset will be the benchmark for a series of future analysis. We intend to obtain spectroscopic follow-up for our brightest candidates, to verify 
our contamination by interlopers. This extended spectroscopic sample, complemented with deep ULTRA-VISTA images in the COSMOS field will be used for 
determining stellar masses, ages, and dust content of faint LBGs at z$\approx$3. It will also be possible to measure the UV slope of galaxies in the 
wavelength range from 1500\AA~to 3000\AA~(rest frame), using a method similar to the one we applied at z$\approx$4 \citep{Cast12}. In addition, based on this sample, 
we will update our measurement of the escape fraction of Ly$\alpha$ continuum, attributed to LBGs at this redshift, in an effort to understand 
their contribution to the reionization of the Universe.

\begin{acknowledgements}
K.B. would like to thank N. Reddy for providing redshifts for the sources in the Q1623 field, which are not publicly available.
We would also like to thank the referee for useful comments that significantly improved the quality of this work. 
\end{acknowledgements}

\bibliographystyle{aa}
\bibliography{biblio}{}

\newpage
\newpage
\begin{appendix}
\section{detailed catalogs for the LBG candidates}
\begin{table*}
\caption{LBG candidates in Q1623 - sample\tablefootmark{*}}         
\centering                   
\begin{tabular}{cccccccccccccc}     
\hline                 
ID & RA & DEC & magR & errR & SN$_R$ & magU & errU & magG & errG & fwhm$_R$ & class$_R$ \\    
\hline                        
  34566 & 246.48277283  &  26.78064728  &    23.13     &  0.01  &   149.69&	 24.36 &      0.03 &   23.18 &      0.01  &    9.36  &     0.03 \\
   5884 & 246.54997253  &  26.61091232  &    23.14     &  0.01  &   210.91&	 25.22 &      0.05 &   23.59 &      0.01  &    5.94  &     0.04 \\
  13708 & 246.49920654  &  26.65563011  &    23.16     &  0.01  &   187.19&	 24.63 &      0.03 &   23.35 &      0.01  &    5.74  &     0.04 \\
  13084 & 246.57254028  &  26.65229034  &    23.17     &  0.01  &   282.61&	 26.38 &      0.09 &   24.17 &      0.01  &    3.80  &     0.98 \\
  21534 & 246.29115295  &  26.86321640  &    23.18     &  0.01  &   211.90&	 27.53 &      0.31 &   24.08 &      0.01  &    3.99  &     0.98 \\
  19912 & 246.32560730  &  26.87409782  &    23.21     &  0.01  &   168.31&	 25.32 &      0.05 &   23.68 &      0.01  &    6.90  &     0.05 \\
  38660 & 246.66195679  &  26.75894356  &    23.21     &  0.01  &   174.62&	 26.20 &      0.11 &   24.10 &      0.02  &    3.99  &     0.98 \\
  32181 & 246.23138428  &  26.79601860  &    23.27     &  0.01  &   177.27&	 24.16 &      0.02 &   23.24 &      0.01  &    5.69  &     0.10 \\
  60060 & 246.45475769  &  26.93926048  &    23.28     &  0.01  &   107.24&	 27.04 &      0.39 &   23.99 &      0.02  &   14.97  &     0.03 \\
  34607 & 246.40966797  &  26.77821350  &    23.29     &  0.02  &    21.34&	 24.87 &      0.27 &   23.46 &      0.07  &   62.89  &     0.00 \\
  57214 & 246.51263428  &  26.97315788  &    23.29     &  0.01  &   139.46&	 27.11 &      0.30 &   24.02 &      0.02  &    4.30  &     0.71 \\
  37109 & 246.31718445  &  26.76618958  &    23.32     &  0.01  &   222.38&	 26.34 &      0.09 &   24.23 &      0.01  &    4.16  &     0.98 \\
  66756 & 246.40708923  &  26.89585304  &    23.32     &  0.01  &   164.64&	 27.07 &      0.24 &   24.07 &      0.02  &    5.40  &     0.20 \\
  53822 & 246.46475220  &  26.99882126  &    23.42     &  0.02  &    81.18&	 26.13 &      0.18 &   24.13 &      0.03  &   10.20  &     0.37 \\
  56256 & 246.59666443  &  26.98033142  &    23.44     &  0.01  &   106.60&	 26.14 &      0.14 &   24.07 &      0.02  &    6.32  &     0.03 \\
   9261 & 246.24270630  &  26.62924004  &    23.44     &  0.01  &   193.67&	 26.24 &      0.09 &   24.27 &      0.02  &    4.40  &     0.94 \\
  62181 & 246.56271362  &  26.93066216  &    23.53     &  0.01  &   173.09&	 26.07 &      0.07 &   24.23 &      0.01  &    3.90  &     0.98 \\
  22697 & 246.44438171  &  26.85557747  &    23.54     &  0.01  &   145.75&	 25.61 &      0.06 &   23.92 &      0.02  &    5.64  &     0.03 \\
  49232 & 246.57498169  &  26.69118500  &    23.54     &  0.01  &   124.11&	-28.87 &      1.08 &   24.07 &      0.02  &    5.36  &     0.05 \\
  68739 & 246.41523743  &  26.87915230  &    23.56     &  0.01  &   104.91&	 26.30 &      0.15 &   23.91 &      0.02  &    6.75  &     0.03 \\
  29424 & 246.36233521  &  26.81218147  &    23.57     &  0.01  &   158.47&	-29.24 &      1.08 &   23.97 &      0.02  &    4.68  &     0.28 \\
  66879 & 246.40129089  &  26.89484978  &    23.58     &  0.02  &   120.78&	 25.32 &      0.05 &   23.89 &      0.02  &    7.51  &     0.03 \\
  35570 & 246.54866028  &  26.77584648  &    23.59     &  0.01  &   118.32&	 25.59 &      0.07 &   24.02 &      0.02  &    7.61  &     0.03 \\
  59234 & 246.48266602  &  26.95467758  &    23.59     &  0.01  &    98.29&	 25.98 &      0.12 &   24.10 &      0.02  &   10.28  &     0.04 \\
  16223 & 246.59768677  &  26.67021179  &    23.60     &  0.01  &   140.01&	 27.54 &      0.36 &   24.56 &      0.02  &    9.00  &     0.03 \\
  23849 & 246.23941040  &  26.84794044  &    23.61     &  0.02  &   116.56&	 27.18 &      0.28 &   24.37 &      0.02  &    5.52  &     0.20 \\
  46293 & 246.40367126  &  26.71342850  &    23.61     &  0.02  &   139.03&	 25.08 &      0.04 &   23.81 &      0.02  &    6.74  &     0.22 \\
\hline 
\end{tabular}
\tablefoot{
\tablefoottext{*}{The full table of all candidates in this field will be available at the CDS.}}
\end{table*}

\begin{table*}
\caption{LBG candidates in COSMOS - sample\tablefootmark{*}}         
\centering                   
\begin{tabular}{cccccccccccccc}     
\hline                 
ID & RA & DEC & magR & errR & SN$_R$ & magU & errU & magG & errG & fwhm$_R$ & class$_R$ \\    
\hline                        
    5479 & 149.96293640 &  2.01424503 &  23.00 &   0.02 &   20.95 &  25.13 &  0.31 &   23.29 &   0.07 &  46.87 &   0.00 \\
   61759 & 149.99604797 &  2.33987594 &  23.02 &   0.02 &   58.57 &  24.58 &  0.06 &   23.19 &   0.03 &  36.24 &   0.00 \\
   32310 & 150.03808594 &  2.21496320 &  23.05 &   0.01 &  117.60 &  27.15 &  0.29 &   23.68 &   0.02 &   9.40 &   0.08 \\
   22408 & 149.85160828 &  2.27638984 &  23.07 &   0.01 &  113.87 &  27.00 &  0.32 &   23.41 &   0.02 &   6.01 &   0.95 \\
   29345 & 149.74142456 &  2.23323703 &  23.07 &   0.02 &   99.14 &  23.96 &  0.03 &   22.95 &   0.02 &   6.21 &   0.05 \\
   46319 & 149.90826416 &  2.12797499 &  23.08 &   0.02 &   40.41 &  24.91 &  0.11 &   23.47 &   0.04 &  31.72 &   0.00 \\
   63930 & 149.89360046 &  2.32787180 &  23.11 &   0.02 &  102.27 & -28.47 &  1.08 &   23.87 &   0.02 &   8.23 &   0.77 \\
   13164 & 149.89323425 &  2.07472181 &  23.15 &   0.01 &  122.98 &  24.97 &  0.04 &   23.20 &   0.02 &   5.43 &   0.28 \\
   21928 & 149.94094849 &  2.27886581 &  23.21 &   0.01 &  101.01 &  25.49 &  0.07 &   23.77 &   0.02 &   7.56 &   0.89 \\
   33783 & 149.98829651 &  2.20698524 &  23.22 &   0.01 &   97.86 &  25.46 &  0.07 &   23.77 &   0.02 &   8.74 &   0.18 \\
   31400 & 150.05088806 &  2.21348929 &  23.24 &   0.02 &   57.07 &  27.32 &  0.60 &   23.79 &   0.03 &   9.59 &   0.04 \\
   45047 & 149.92820740 &  2.13668013 &  23.24 &   0.01 &  109.93 &  26.00 &  0.10 &   23.59 &   0.02 &   5.10 &   0.69 \\
   32549 & 150.03649902 &  2.21406007 &  23.25 &   0.02 &   66.00 &  26.75 &  0.31 &   23.76 &   0.03 &  21.06 &   0.00 \\
   13356 & 149.80078125 &  2.07625198 &  23.30 &   0.02 &   79.69 &  25.14 &  0.07 &   23.68 &   0.03 &   8.78 &   0.02 \\
   35450 & 149.86567688 &  2.19670677 &  23.31 &   0.02 &   40.35 &  25.05 &  0.10 &   23.60 &   0.04 &  31.72 &   0.00 \\
   53388 & 150.08761597 &  2.39437175 &  23.31 &   0.02 &   69.07 &  25.16 &  0.07 &   23.58 &   0.03 &   6.04 &   0.03 \\
   40101 & 149.83178711 &  2.16793180 &  23.33 &   0.01 &   92.38 &  26.54 &  0.19 &   23.80 &   0.02 &   7.80 &   0.03 \\
   36886 & 149.74243164 &  2.18706799 &  23.35 &   0.02 &   83.28 &  24.33 &  0.04 &   23.23 &   0.02 &   5.01 &   0.44 \\
   31631 & 150.03479004 &  2.21980000 &  23.38 &   0.02 &   99.90 &  25.43 &  0.06 &   23.83 &   0.02 &   5.82 &   0.09 \\
   12807 & 149.81239319 &  2.06983995 &  23.40 &   0.02 &   46.72 &  26.09 &  0.24 &   24.15 &   0.04 &  63.81 &   0.00 \\
\hline 
\end{tabular}
\tablefoot{
\tablefoottext{*}{The full table of all candidates in this field will be available at the CDS.}}
\end{table*}

\begin{table*}
\caption{LBG candidates in Q0933 - sample\tablefootmark{*}}         
\centering                   
\begin{tabular}{cccccccccccccc}     
\hline                 
ID & RA & DEC & magR & errR & SN$_R$ & magU & errU & magG & errG & fwhm$_R$ & class$_R$ \\    
\hline                        
 20617  &   143.34263611  &    28.64185143 &  23.01 &	0.02  &  71.39  &  26.00 &  0.12 &  23.84  &  0.03  &	10.87 &   0.03  \\ 
 31260  &   143.47781372  &    28.53086853 &  23.02 &	0.02  &  93.00  &  24.38 &  0.03 &  23.14  &  0.02  &	 4.97 &   0.98  \\ 
  9982  &   143.52899170  &    28.74859619 &  23.04 &	0.02  &  87.35  &  25.06 &  0.04 &  23.34  &  0.02  &	 5.28 &   0.98  \\ 
 21462  &   143.34049988  &    28.63321877 &  23.05 &	0.02  &  66.98  &  24.62 &  0.04 &  23.14  &  0.03  &	10.19 &   0.56  \\ 
  6971  &   143.32162476  &    28.78584099 &  23.07 &	0.02  &  32.81  &  25.55 &  0.15 &  23.69  &  0.05  &	34.86 &   0.02  \\ 
 23901  &   143.52601624  &    28.60801506 &  23.14 &	0.02  &  80.30  &  24.57 &  0.03 &  23.34  &  0.02  &	 5.23 &   0.89  \\ 
 16547  &   143.24850464  &    28.68159676 &  23.17 &	0.02  &  94.24  &  26.72 &  0.16 &  23.71  &  0.03  &	 5.93 &   0.08  \\ 
 24781  &   143.42724609  &    28.59820366 &  23.23 &	0.02  &  82.02  &  24.43 &  0.03 &  23.23  &  0.03  &	 6.13 &   0.40  \\ 
  6902  &   143.45941162  &    28.78800392 &  23.23 &	0.02  &  70.67  &  25.56 &  0.05 &  23.60  &  0.03  &	 6.84 &   0.15  \\ 
 21093  &   143.61053467  &    28.63752556 &  23.25 &	0.03  &  56.01  &  24.92 &  0.05 &  23.56  &  0.04  &	 7.61 &   0.06  \\ 
 32628  &   143.38154602  &    28.51601219 &  23.27 &	0.02  &  55.21  &  25.38 &  0.07 &  23.77  &  0.03  &	10.80 &   0.03  \\ 
 33782  &   143.57832336  &    28.50344086 &  23.33 &	0.02  &  91.28  &  26.10 &  0.10 &  23.72  &  0.02  &	 4.97 &   0.98  \\ 
 30715  &   143.45211792  &    28.53575134 &  23.35 &	0.05  &  36.56  & -28.48 &  1.08 &  24.35  &  0.06  &	12.42 &   0.02  \\ 
 24839  &   143.26560974  &    28.59762192 &  23.36 &	0.02  &  78.93  &  26.50 &  0.12 &  24.32  &  0.03  &	 5.55 &   0.97  \\ 
 10354  &   143.28079224  &    28.74371147 &  23.40 &	0.02  &  66.90  &  26.56 &  0.13 &  24.37  &  0.03  &	 6.36 &   0.17  \\ 
 31459  &   143.60533142  &    28.52896500 &  23.40 &	0.03  &  64.49  &  26.34 &  0.12 &  24.22  &  0.03  &	 4.83 &   0.98  \\ 
 20655  &   143.27740479  &    28.64189720 &  23.41 &	0.03  &  54.94  &  25.22 &  0.06 &  23.72  &  0.04  &	11.97 &   0.03  \\ 
\hline 
\end{tabular}
\tablefoot{
\tablefoottext{*}{The full table of all candidates in this field will be available at the CDS.}}
\end{table*}

\end{appendix}

\end{document}